\documentstyle[11pt,aaspp4]{article}

\def\lea{\mathrel{<\kern-1.0em\lower0.9ex\hbox{$\sim$}}}
\def\gea{\mathrel{>\kern-1.0em\lower0.9ex\hbox{$\sim$}}}

\slugcomment{To appear in The Astronomical Journal, August 1998}

\lefthead{Sarajedini}
\righthead{Clusters in the LMC Age Gap}

\begin{document}

\def\et{\hbox{et~al.\ }}

\def\feh{\hbox{[Fe/H]}}
\def\ofe{\hbox{[O/Fe]}}
\def\deg{\hbox{$^\circ$}}
\def\sun{\hbox{$\odot$}}
\def\earth{\hbox{$\oplus$}}
\def\lesssim{\mathrel{\hbox{\rlap{\hbox{\lower4pt\hbox{$\sim$}}}\hbox{$<$}}}}
\def\gtrsim{\mathrel{\hbox{\rlap{\hbox{\lower4pt\hbox{$\sim$}}}\hbox{$>$}}}}
\def\sq{\hbox{\rlap{$\sqcap$}$\sqcup$}}
\def\arcmin{\hbox{$^\prime$}}
\def\arcsec{\hbox{$^{\prime\prime}$}}
\def\fd{\hbox{$.\!\!^{\rm d}$}}
\def\fh{\hbox{$.\!\!^{\rm h}$}}
\def\fm{\hbox{$.\!\!^{\rm m}$}}
\def\fs{\hbox{$.\!\!^{\rm s}$}}
\def\fdg{\hbox{$.\!\!^\circ$}}
\def\farcm{\hbox{$.\mkern-4mu^\prime$}}
\def\farcs{\hbox{$.\!\!^{\prime\prime}$}}
\def\fp{\hbox{$.\!\!^{\scriptscriptstyle\rm p}$}}
\def\micron{\hbox{$\mu$m}}
 
\let\la=\lesssim
\let\ga=\gtrsim

\title{Three Populous Clusters Discovered to be in the LMC 
Age Gap\altaffilmark{1}}

\author{Ata Sarajedini\altaffilmark{2,}\altaffilmark{3}}
\affil{San Francisco State University, Department of Physics and
Astronomy \\ 1600 Holloway Avenue, San Francisco, CA 94132 \\ {\it ata@stars.sfsu.edu}}

\altaffiltext{1}{Based on observations made with the NASA/ESA Hubble
Space Telescope obtained at the Space Telescope Science Institute,
which is operated by the Association of Universities for Research in
Astronomy, Incorporated, under NASA contract NAS5-26555.}

\altaffiltext{2}{Hubble Fellow}

\altaffiltext{3}{Guest User, Canadian Astronomy Data Centre, which is
operated by the Dominion Astrophysical Observatory for the National
Research Council of Canada's Herzberg Institute of Astrophysics.}


\begin{abstract}

In the process of searching the Hubble Space Telescope archive, we 
have serendipitously discovered three populous Large Magellanic Cloud 
(LMC) clusters with ages that place them in the LMC `age gap.'. These clusters
- NGC 2155, SL663, and NGC 2121 - turn out to
have $[Fe/H]$$\sim$ --1.0 and ages of $\sim$4 Gyr. This puts them
in the age gap between the intermediate-age LMC clusters, the oldest of 
which are $\sim$2.5 Gyr old, and ESO121-SC03, which has an age of 
$\sim$ 9 Gyr. The addition of these three
clusters to the LMC age - metallicity relation has reduced the
discrepancy between the age distribution of the LMC clusters and the field
stars. Furthermore, it indicates that searches
to find more clusters older than $\sim$2.5 Gyr in the LMC are crucial
to a better understanding of its global star formation history.

\end{abstract}



%

\section{Introduction}

The chemical enrichment/star formation history (SFH) is an 
identifying feature of 
every self-gravitating stellar system. One manifestation of this is the 
relation between
age and metallicity among the star clusters in a given galaxy. Empirical
information on how cluster age and metal abundance correlate provides an
important clue that will eventually allow us to understand how star formation
(and hence chemical enrichment) proceeds under a variety of potentially
influential circumstances. It is for this reason that we strive to
better define the age - metallicity relations of the cluster and field
populations in galaxies. Nearby
galaxies are no exception, especially the Large and Small Magellanic
Clouds (L/SMC) which have provided numerous puzzles and challenges
for theorists. One of the most persistent of these has been the `age gap' 
seen between the intermediate-age clusters ($t \lea 2.5$ Gyr;
see Sec. 4.1 for a justification of this limit) and the
old clusters ($t \gea 13$ Gyr) in the LMC (Geisler et al. \cite{geisea1997}). 
The LMC age gap is also a metallicity gap in the sense that the 
intermediate-age clusters
have $\langle$$[Fe/H]$$\rangle$$\sim$$-0.5$, while the old clusters
are closer to $\langle$$[Fe/H]$$\rangle$$\sim$$-2$.
There is only one cluster that is known to lie in the gap  - ESO121-SC03 with 
an age of
$\sim$9 Gyr and $[Fe/H]\sim-1$. In contrast, recent studies based
on Hubble Space Telescope (HST) data of the LMC field stars
tell a signficantly different story (see Geha et al. \cite{gehea1998} for
a review). While there are variations that depend 
on position in the LMC (Vallenari et al. \cite{valea1996}), the 
global star formation rate has been fairly constant
for {\it most} of the LMC's history. However, sometime between 2 
and 4 Gyr ago, a
burst of star formation occurred producing the present population of
young to intermediate age field stars and clusters (Bertelli et al.
\cite{bertea1992}; Vallenari et al. \cite{valea1996}; 
Gallagher et al. \cite{gallea1996}; Holtzman et al.
\cite{holtzea1997}). Furthermore, the models produced by 
Geha et al. (\cite{gehea1998}) suggest that roughly half of the LMC
field stars are older than 4 Gyr. What this implies is
that there should be many star clusters with ages between $\sim$2.5 Gyr and 
$\sim$13 Gyr, which were
formed contemporaneously with the field stars. The obvious question of
course is: where are these `age gap' clusters?

A number of investigators have undertaken surveys to find clusters in
the LMC age gap (Da Costa \cite{dac1991}; Geisler et al. \cite{geisea1997},
and references therein). The overwhelming conclusion has been that the
LMC contains only one cluster with an age between 2.5 Gyr and 13 Gyr
(i.e. ESO121-SC03). There is the possibility that some clusters did
exist in the age gap sometime in the past, and that these have either
dissolved into the LMC field or been stripped off (Olszewski \cite{olsz1993}).
However, it is difficult to see how this process can preferentially affect only
clusters formed during a given epoch, unless the mass spectrum of 
density fluctuations
producing the gap clusters was somehow different from that which produced
the other clusters in the LMC. The possible reasons for this are not obvious.


While understanding the SFH of the LMC is extremely important, this
was not our original aim when we began this study. Initially, we were
searching the Hubble Space Telescope (HST) archive looking for intermediate
age clusters that display the red giant branch (RGB) Bump in their
color-magnitude diagrams (CMD). We did find one such cluster (NGC 411; see
Alves \& Sarajedini \cite{alsa1998}). However, more importantly for the
present work, we discovered
three clusters whose CMDs exhibit significant numbers of stars in the
Hertzsprung gap, indicative of ages older than $\sim$2.5 Gyr. These clusters
are NGC 2155, NGC 2121, and SL663, and the next section describes 
the observations of these clusters and the data reduction. Section 3 
presents the CMDs while Section 4 discusses
the ages yielded by these CMDs. The implications of these results are
detailed in Section 5.

\section{Observations and Reductions}

The observations used in the present study were taken with the
Hubble Space Telescope (HST) Wide Field Planetary Camera 2 (WFPC2)
during the period 1994 January 30 to February 2. For each cluster,
one 230s exposure and one 120s exposure were obtained in the F450W 
($\approx$B) and F555W ($\approx$V) filters, respectively, with the gain
value set at 14 e/ADU. The telescope
was not moved between these observations. The images were recalibrated
at the Canadian Astronomy Data Centre using the `recommended'
calibration frames. The frames were then multiplied by the geometric
distortion correction frame (Biretta \cite{bir1995}). Before beginning the
photometric reduction process, the bad pixel masks were applied to
the program frames thereby flagging pixels known to be defective.
We note in passing that the identification of one of the clusters,
SL 663, appears to be incorrect in the image header. The position of
the target ($\alpha$$_{J2000}$ = 5$^h$ 42$^m$ 30$^s$,
$\delta$$_{J2000}$ = --65$\deg$ 21' 49'')
is coincident with the cluster SL 663 in the 
Kontizas et al. (\cite{konea1990}) catalogue not SL 633 as suggested
by the image header (which is the 30 Doradus cluster). Table 1 gives 
the Journal of the observations.

Aperture photometry was performed using the routines in DAOPHOT II
(Stetson \cite{st1994}). A nominal aperture radius of 2 pixels was used with
aperture corrections providing the offset to total magnitudes inside
of a 0.5 arcsec radius. The typical errors in the aperture corrections
were $\pm$0.05 mag. After matching the instrumental magnitudes to form
colors, the following equations were used to transform our data to
the standard system defined by Holtzman et al. (\cite{holtz1995}):
\begin{eqnarray}
B=b+a_B+b_B(B-V)+c_B(B-V)^2
\end{eqnarray}
\begin{eqnarray}
V=v+a_V+b_V(B-V)+c_V(B-V)^2, 
\end{eqnarray}
where $b$ and $v$ refer to the instrumental magnitudes
(i.e. aperture magnitude + aperture correction + 2.5Log(t) -- 25.0), and
$B$, $V$, and $(B-V)$ are the standard system values. The 
coefficients for F450W and F555W are taken from Table 10 and 7
of Holtzman et al. (\cite{holtz1995}), respectively. A 12\% CTE correction 
in the y-axis direction was also included (Holtzman et al. \cite{holtz1995}) 
because these data were 
taken at an operating temperature of --76$\deg$ C before WFPC2 was cooled to
--88$\deg$ C.

\section{Color-Magnitude Diagrams}

We first seek to investigate the integrity of our data reduction method.
To accomplish this, we have chosen the LMC cluster NGC 2193, which 
Geisler et al. ({\cite{geisea1997}) claim has an age of 2.3 Gyr, one of
the oldest among the intermediate age clusters (IACs). Figure 1 shows the
$B-V$ color-magnitude diagram (CMD) of NGC 2193 from archival HST
observations made in the F450W and F555W passbands. Only the Planetary Camera
data are plotted. In the same figure,
we have included the ground-based photometry of NGC 2193 from the
work of Da Costa et al. (\cite{dakimo1987}). These latter data were
originally observed in $B-R$ and have been transformed to $B-V$ using
the well-determined relation published by Mighell et al. (\cite{misafr1998a}).
The agreement between these datasets is striking. All of the CMD
features appear to coincide between these two datasets. This agreement
lends credence to our reduction methods and lays the foundation for
the remainder of the paper. 

The lower panel of Fig. 1 compares the CMD of NGC 2193 to that of
the cluster SL 556 (Hodge 4),
to which Geisler et al. ({\cite{geisea1997}) assign an age of $\sim$3 Gyr.
This is the oldest cluster among the IAC according to Geisler et al. 
({\cite{geisea1997}), even older than NGC 2193 based on their analysis.
However, the clear correspondence of the HST data for NGC 2193 and
Hodge 4 indicates that in fact these two clusters are more likely to be
the same age. We return to this issue in the next section.

Figures 2 through 4 illustrate the CMDs for the program clusters in
this study - NGC 2155, SL 663, and NGC 2121, respectively - the clusters 
that appear to lie in the LMC age gap (see below). In each
figure, CMDs for the four WFPC2 CCD chips (PC1, WF2, WF3, and WF4) are shown. 
The clusters were centered on the PC1 aperture so
that this CMD exhibits the cluster principal sequences
most clearly. The strongest case is that of NGC 2155 which displays
a blue hook near the main sequence turnoff (MSTO) and a well-developed
subgiant branch indicative of an older population. The CMD of SL 663
also displays these features, but because SL 663 is not as populous as
NGC 2155, the features are not as obvious. The morphology of the NGC 2121
main sequence turnoff is similar to that of NGC 2155, but
its subgiant branch is not as well developed.

\section{Cluster Ages}

\subsection{Comparison With Theoretical Isochrones}

One method we can utilize to estimate the cluster ages is simply to 
overplot isochrones on the CMDs of the clusters presented herein. 
We will make use of the tracks published
by Bertelli et al. (\cite{be1994}), which are given for Z=0.001 
($[Fe/H] = -1.3$) and Z=0.004 ($[Fe/H] = -0.7$) and include
the effects of convective core overshooting for stars
more massive than one solar mass. Before, we proceed with the isochrone
comparisons, we 
need an estimate for the metallicity of each cluster. In the case of 
NGC 2193 and Hodge 4, we look to the
metallicity values derived from published CMD photometric studies. 
In particular,
Mateo \& Hodge (\cite{maho1986}) derive $[Fe/H] = -0.7\pm0.3$ for 
Hodge 4 and Da Costa et al. (\cite{dakimo1987}) find
$[Fe/H] = -0.5\pm0.3$ for NGC 2193. For the sake of convenience in
the isochrone comparison, we adopt $[Fe/H] = -0.7$ for both of
these clusters. 

Our preferred metallicity indicator for NGC 2155, SL 633, and NGC 2121 is 
the slope of the RGB 
because it is independent of reddening and uncertainties in the
photometric zeropoint. The latter could be significant given the
fact that the data have been derived from HST WFPC2 observations taken
before the cooldown to --88$\deg$ C. In addition, Mighell et al.
(\cite{misafr1998b}) have shown that, for clusters as young as 4 Gyr,
the RGB slope metallicities are only 
$0.11\pm0.06$ dex more metal-poor as compared with metal abundances
estimated from the Calcium triplet (Da Costa \& Hatzidimitriou 
\cite{daha1998}). We note in passing that this method is not applicable to
NGC 2193 and Hodge 4 because these clusters are younger than $\sim$4 Gyr,
whereas we show below that the program clusters are $\sim$4 Gyr old.
To facilitate the measurement of the RGB slope,
we begin by fitting a polynomial to the RGB stars using the iterative
2$\sigma$ rejection procedure described by Sarajedini \& Norris 
(\cite{sano1994}) applied to the combined photometry from all four
WFPC2 CCDs for each cluster. These fits are shown in Fig. 5. Utilizing 
the relation published by 
Mighell et al. (\cite{misafr1998b}) for the RGB slope measured 2 magnitudes
above the HB, we find $[Fe/H] = -1.08 \pm 0.12$ for NGC 2155,
$[Fe/H] = -1.05 \pm 0.16$ for SL 663, and $[Fe/H] = -1.04 \pm 0.13$ for 
NGC 2121. Thus, all three of these clusters share approximately the 
same metallicity, which we take to be $[Fe/H] = -1.0$ for the isochrone
comparison. These low metallicities are also a fairly strong indicator of 
relatively old age
(see the age-metallicity relation of Olszewski et al. \cite{olszea1991}),
although there are some IACs with comparably low metallicities 
(Bica et al. \cite{bicaea1998}).

Figures 6 through 9 show our Planetary Camera photometry for the 5 
LMC clusters of this paper and the theoretical isochrones of 
Bertelli et al. (\cite{be1994}). For the sake of completeness,
Fig, 10 displays the isochrone
fit to the CCD photometry of ESO121-SC03 from 
Mateo et al. (\cite{mahosc1986}). In each figure, the isochrones 
have been shifted vertically by matching the magnitude of the
red HB clump and horizontally by matching the location of the 
unevolved main
sequence {\it below} the turnoff. This approach eliminates the need to rely
on the integrity of the photometric zeropoints, the reddening, and the distance
to the LMC as well as reducing the influence of various theoretical 
uncertainties in the models
(e.g. mixing length parameter, helium diffusion, opacities, etc.)

The two panels of Fig. 6 illustrate the isochrone comparisons to
the photometry of NGC 2193 and Hodge 4. These clusters are believed to be
among the oldest IAC; as such, knowledge of their precise ages is 
important in setting the upper age limit of the IAC. From the 
appearance of Fig. 6, we derive an age of $2.0 \pm 0.2$ Gyr for 
NGC 2193 and Hodge 4.
The quoted error is our best estimate of the 1$\sigma$ uncertainty
inherent in this age determination. The result that both NGC 2193 and 
Hodge 4 have similar ages is not surprising given the comparison shown 
in the lower panel of Fig. 1. In contrast to these results, the work of 
Geisler et al. (\cite{geisea1997}) indicates that their oldest IAC (Hodge 4) 
has an age of $\sim$3 Gyr, while NGC 2193 is given an age of 2.3 Gyr. 
The higher-quality HST data presented herein has allowed us to revise these 
ages downward somewhat.  
Our upper age limit for the IAC agrees with that of Bomans et al. 
(\cite{bomea1995}), who performed isochrone fits to 17 such clusters. 
They find that the oldest cluster in their sample is NGC 1978 with an 
age of 2.2 Gyr. Hodge 4 and NGC 2193 are assigned ages of 2.0 Gyr and 
1.6 Gyr, respectively. Taken together, the evidence suggests that
2.2 Gyr is a good estimate for the age of the oldest IAC. However,
to be conservative and keeping in mind the uncertainty in these ages, we adopt
2.5 Gyr as the boundary between the age of the oldest IAC and the
onset of the age gap.

The two panels of Figs. 7, 8, and 9 show the isochrone comparisons to
NGC 2155, SL 633, and NGC 2121, respectively. In all cases, the Z=0.001 
isochrones are displayed in the upper panel, while the Z=0.004 isochrones
are included in the lower panel. Recall that our metallicity determination
indicated that $[Fe/H]\sim-1.0$ for these clusters. 
As a result, we show comparisons to both metallicities with the 
intention of interpolating between them. Based on the position
of the main sequence turnoff (MSTO) and the subgiant branch, we infer an 
age of 4.5 Gyr from the Z=0.001 isochrones and 3.5 Gyr from the
Z=0.004 isochrones. Taking the mean of these values, we conclude that 
NGC 2155, SL 633, and NGC 2121 all have ages of $4.0\pm0.3$, where, again
we have attempted to estimate the 1$\sigma$ uncertainty. As a result,
it appears as though these three clusters have ages which
are older than the oldest known IAC; this places them squarely in the 
LMC age gap. We point out however that if the metal abundances of these
clusters are in reality higher than we have estimated herein, their
ages will be correspondingly younger, as evidenced by the Z=0.004 isochrone
fits. We return to the implications of this possibility in Sec. 5.

We now seek to establish the
age of ESO121-SC03 based on the same techniques utilized above for
the other clusters. From the slope of the cluster RGB measured from the
RGB fit calculated by Sarajedini et al. (\cite{salele1995}), we find
$[Fe/H] = -1.01 \pm 0.15$. This is in good agreement with 
$[Fe/H] = -0.93 \pm 0.1$ determined by Olszewski et al. (\cite{olszea1991}) 
from Calcium triplet spectroscopy of 3 stars. Thus, we 
compare the photometry of ESO121-SC03 to isochrones with abundances that
bracket this value and interpolate between them to estimate the cluster age.
Figure 10 shows such a comparison; from the appearance of this figure,
we infer an age of 11 Gyr from the Z=0.001 isochrones and 8 Gyr from
the Z=0.004 isochrones. Taking the mean of these values results in an
age of $9.5\pm 0.5$ Gyr (estimated 1$\sigma$ error). This is in
excellent agreement with the results of several previous investigators
(e.g. Geisler et al. \cite{geisea1997}; Mateo et al. \cite{mahosc1986}). 


\subsection{$\delta$$V$ Ages}

Another method at our disposal is to estimate the cluster ages via
morphological indicators in the CMD as formulated by Phelps et al. 
(\cite{phelea1994}). They define the `MSTO' magnitude as the point
midway between the
bluest point of the actual MSTO and the base of the RGB. The difference in
magnitude between this point and the red HB clump is designated by
$\delta$$V$. They also define the difference in color ($\delta$$(B-V)1$) 
between the MSTO and the point on the lower RGB located one magnitude above
the MSTO. Using an empirically derived equation 
($\delta$$V = 3.77 - 3.75\times$$\delta$$(B-V)1$), they convert 
the measured $\delta$$(B-V)1$ values to $\delta$$V$. The resulting quantity is 
combined with $\delta$$V$ to create a mean morphological age indicator
designated by $\langle$$\delta$$V\rangle$. 

Following the methods described 
by Phelps et al. (\cite{phelea1994}), we have computed 
$\langle$$\delta$$V\rangle$ for
the 6 LMC clusters in this paper. The results are shown in
Table 2. Janes \& Phelps (\cite{japh1994}) discuss the use of this
$\langle$$\delta$$V\rangle$ parameter to estimate ages. 
They combine $\langle$$\delta$$V\rangle$
values for 26 open clusters and 39 Galactic globular clusters to
parameterize age as a function of $\langle$$\delta$$V\rangle$ 
(Age = 0.73$\times10^{(0.256\langle}$$^{\delta}$$^{V\rangle + 0.0662\langle}$$^{\delta}$$^{V\rangle^2}$). 
We have applied this
relationship to our $\langle$$\delta$$V\rangle$ values to arrive at 
the ages listed
in column 4 of Table 3. Janes \& Phelps (\cite{japh1994}) caution that their
formulation is most reliable when used in a relative sense and should
not be trusted to yield precise absolute ages. Thus, it is of interest
to investigate the offset between the $\langle$$\delta$$V\rangle$ ages 
and those
estimated from isochrone fitting. Column 5 of Table 2 lists these
age differences. The mean
of the differences is $1.05 \pm 0.09$ Gyr. The small error associated
with this value suggests that the relative ages as derived from
the two methods are highly robust. Therefore, because the 
$\langle$$\delta$$V\rangle$ ages
have an uncertainty in their zeropoint, for the remainder of this paper,
we adopt the cluster ages as given by the isochrone comparisons.

\subsection{Comparison With Previous Results}

We begin by noting that there is very little previous work on the
properties of the cluster SL663. Elson \& Fall (\cite{elfa1988})
quote an age of $2.2 \pm 0.4$ Gyr based on an unpublished CMD from
Mario Mateo. Aside from its position on the sky, no other information 
exists in the literature for SL663.

The situation is marginally better for the cluster NGC 2155.
Searle et al. (\cite{sewiba1980}) classify it as an SWB type VI
cluster based on its location in the integrated light Q(ugr) vs Q(vgr) 
diagram. Elson \& Fall (\cite{elfa1988}), again
quoting an unpublished CMD from Mario Mateo, list an age of
$2.5 \pm 0.6$ Gyr. Olszewski et al. (\cite{olszea1996}) prefer an
age of 3.5 Gyr for NGC 2155. The sole published CMD for NGC 2155 is that of
Hesser et al. (\cite{hehaug1976}) which is based on photographic
plates and is not transformed to magnitudes on a standard photometric
system. As a result, the current utility of this diagram is questionable.
Metallicity measurements for NGC 2155 range from $[Fe/H] = -0.55$
(Olszewski et al. \cite{olszea1991}) to $[Fe/H] = -1.2\pm0.2$
(Bica et al. \cite{bicaea1986}), although the former has generally
been preferred in previous work since it is based on Calcium triplet 
spectroscopy of individual stars (albeit only 3) whereas the latter is 
based on integrated DDO photometry. We return to this point below.

Lastly, NGC 2121 has received considerable attention in previous studies. 
Like NGC 2155, Searle et al. (\cite{sewiba1980}) classified NGC 2121
as an SWB type VI cluster. The photographic CMD presented by
Flower et al. (\cite{flowea1983}) reveals a cluster with an age of
2.0 Gyr (Geisler et al. \cite{geisea1997}).
Metallicity measurements for NGC 2121 include $[Fe/H] = -0.95 \pm 0.4$
(Cohen \cite{coh1982}; Fe, Ca, Na, and Mg line widths for two stars),
$[Fe/H] = -0.75\pm0.25$ (Bica et al. \cite{bicaea1986}; integrated DDO 
photometry), $[Fe/H] = -0.61$ (Olszewski et al. \cite{olszea1991}; Calcium 
triplet spectra for two stars), and $[Fe/H] = -0.10\pm0.21$ 
(de Freitas Pacheco et al. \cite{defrea1998}; integrated-light
spectral indices).  

The Calcium triplet metallicities published by 
Olszewski et al. (\cite{olszea1991}) have received general acceptance
in the literature. They are thought to be relatively reliable. However, 
Bica et al. ({\cite{bicaea1998}) have found that the metallicities published by
Olszewski et al. (\cite{olszea1991}) for IACs of similar age to their sample
are more metal-rich by $\sim$0.25 dex. The nature of this difference is 
unknown and could be due to a metallicity gradient in the LMC disk 
(since the Bica et al. clusters are generally further out) or to a 
difference in the metallicity values themselves. Adding support
to the latter possibility is the fact that for NGC 2155 and NGC 2121, our 
RGB slope analysis yields metallicities that are $\sim$0.4 dex 
more metal-poor than those values quoted by 
Olszewski et al. (\cite{olszea1991}).

It is clear from the above discussion that the published values for the 
ages and metallicities of SL663, NGC 2155, and NGC 2121 span a large range.
In the present paper, we have combined HST/WFPC2 photometry for 5 clusters
along with ground-based CCD photometry for another to establish a
robust relative age and abundance ranking for these clusters.
Because these ages are based on the location of
the main sequence turnoff, they should be considered the most reliable
yet determined for these clusters. In addition, even though our
metallicity measurements are not based on spectra for a large number of
individual stars in each cluster, they are also likely to be the
most reliable to date because our method has been tested previously
on SMC clusters (Mighell et al. \cite{misafr1998b}).

\section{Discussion and Conclusions}

The resulting relation between age and metallicity for the LMC star
clusters is shown in Fig. 11. The open symbols are the ages and 
abundances from the 
work of Geisler et al. (\cite{geisea1997}) supplemented by additional
clusters from Bica et al. (\cite{bicaea1998}) and the values for
NGC 2193, Hodge 4, and ESO121-SC03 from this paper. The filled square
represents the location of our three `age gap' clusters. Clearly,
the clusters NGC 2155, NGC 2121, and SL 663 do indeed fall in the
age gap between $\sim$2.5 Gyr and $\sim$13 Gyr. The reader should
keep in mind, however, that if the metallicities of these clusters
are higher than the values we have adopted herein, their ages will
be correspondingly younger (see Sec. 4.1). As such, they may
eventually be considered as belonging to the old-age tail of the 
IAC distribution. Future spectroscopic abundance measurements will
shed more light on this.

In any event, the addition of these three clusters to the age - metallicity
relation of the LMC has not eliminated the discrepancy between the
cluster age distribution and that of the field stars.
If there are no more clusters to be discovered in the gap, then
we will require some explanation for why the clusters and the field
stars exhibit such differing SFHs.
However, what is {\it more} likely is that there are as yet unstudied
clusters in the LMC that will further fill in the age gap. Future
ground-based and HST photometric surveys may reveal more such clusters.

For the present paper, we have utilized archival HST/WFPC2 images of 
LMC populous
clusters to show that there are at least three clusters in the LMC
age gap - NGC 2155, SL663, and NGC 2121. These clusters have
$[Fe/H]$$\sim$ --1.0 and ages of $\sim$4 Gyr. The addition of these three
clusters to the LMC age - metallicity relation is the first step in
reducing the significant difference between the inferred SFHs of the LMC 
clusters and the field stars. This strongly indicates that searches
to find more clusters older than $\sim$2.5 Gyr in the LMC are crucial
to a better understanding of its global SFH.


%
%

\acknowledgments

We acknowledge fruitful conversations with Eva Grebel, Doug Geisler,
and Pierre Demarque.
We are grateful to Eva Grebel and Gary Da Costa for a careful reading
of this manuscript as well as the referee, Doug Geisler, whose comments 
greatly improved the quality of this work.
Ata Sarajedini would like to express his gratitude to UCO/Lick Obs.
for kind hospitality during his visit. Ata Sarajedini was supported by the National Aeronautics and Space 
Administration (NASA) grant number HF-01077.01-94A from
the Space Telescope Science
Institute, which is operated by the Association of Universities for
Research in Astronomy, Inc., under NASA contract NAS5-26555.

\newpage
 
%
%
\figcaption[fig1.eps]{
\label{fig-1}
The upper panel shows the color-magnitude diagram for NGC 2193 from 
HST/WFPC2 from the present work (open circles)
and the ground-based study of Da Costa et al. (1987; X's). 
The latter have been converted from $B-R$ to $B-V$ using the equation
derived by Mighell et al. (1998a). Note the good agreement between the 
two datasets. The lower panel compares the NGC 2193 photometry from
the upper panel (open circles) with our HST/WFPC2 photometry for
SL 556 (Hodge 4; X's). The latter has been shifted by +0.05 in B--V
to match the location of the NGC 2193 red HB clump. Note that these two 
clusters are almost identical in age (main sequence turnoffs coincide)
and metallicity (red giant branches coincide).}
\figcaption[fig2.eps]{
\label{fig-2}
Color-magnitude diagrams for NGC 2155 for each individual WFPC2 CCD chip.}
\figcaption[fig3.eps]{
\label{fig-3}
Color-magnitude diagrams for SL 663 for each individual WFPC2 CCD chip.}
\figcaption[fig4.eps]{
\label{fig-4}
Color-magnitude diagrams for NGC 2121 for each individual WFPC2 CCD chip.}
\figcaption[fig5.eps]{
\label{fig-5}
The red giant branch (RGB) polynomial fits used to estimate the slope
of the RGB, which is then utilized to measure the cluster metallicity.}
\figcaption[fig6.eps]{
\label{fig-6}
Comparison of theoretical isochrones with the Planetary Camera
photometry for NGC 2193
and Hodge 4. We have adopted a metallicity of [Fe/H] = --0.7 for these
clusters. The isochrones have been shifted vertically by matching 
the magnitude of the red HB clump and horizontally by matching the 
location of the unevolved main sequence {\it below} the turnoff.
The isochrone fits yield an age of 2.0 Gyr for both of
these clusters.}
\figcaption[fig7.eps]{
\label{fig-7}
Same as Fig. 6 except that we plot the Planetary Camera photometry
for NGC 2155 and include isochrones for two metallicity values that bracket the 
abundance of the cluster. A mean age of $4.0\pm0.3$ Gyr is indicated by 
these comparisons.}
%
\figcaption[fig8.eps]{
\label{fig-8}
Same as Fig. 7 except that the Planetary Camera photometry
for SL 663 is shown.}
\figcaption[fig9.eps]{
\label{fig-9}
Same as Fig. 7 except that the Planetary Camera photometry
for NGC 2121 is shown.}
\figcaption[fig10.eps]{
\label{fig-10}
Same as Fig. 7 except that the ground-based CCD photometry for
ESO121-SC03 (Mateo et al. 1986) is shown.  A mean age 
of $9.5\pm0.5$ Gyr is indicated by these comparisons.}
\figcaption[fig11.eps]{
\label{fig-11}
The relation between age and metallicity for LMC populous
clusters. The open circles are data from Geisler et al. (1997) and
Bica et al. (1998) supplemented by our determinations for NGC 2193, Hodge 4,
and ESO121-SC03, while the filled square represents the location of
the three `age gap' clusters, NGC 2155, SL 663, and NGC 2121.}

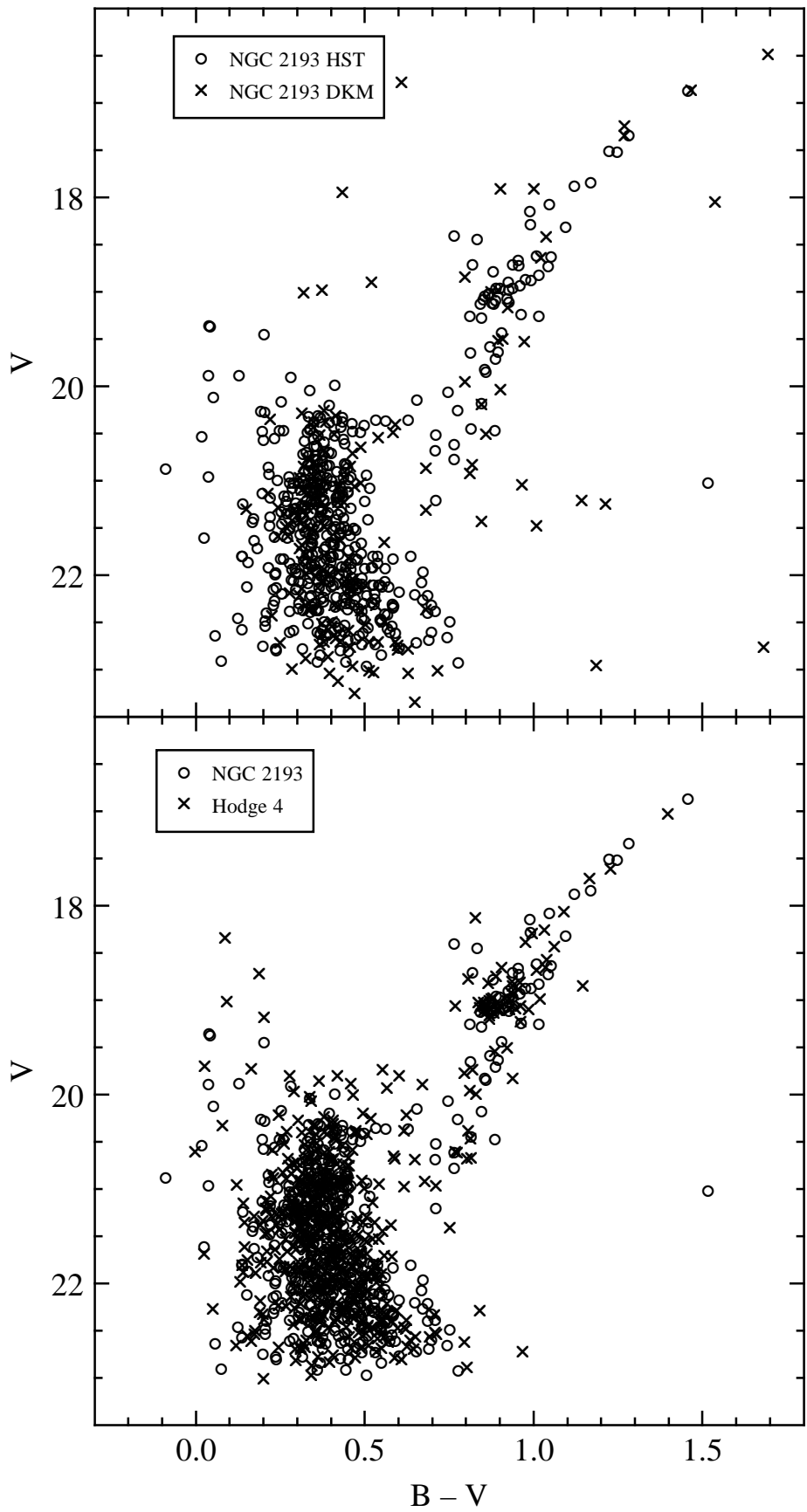
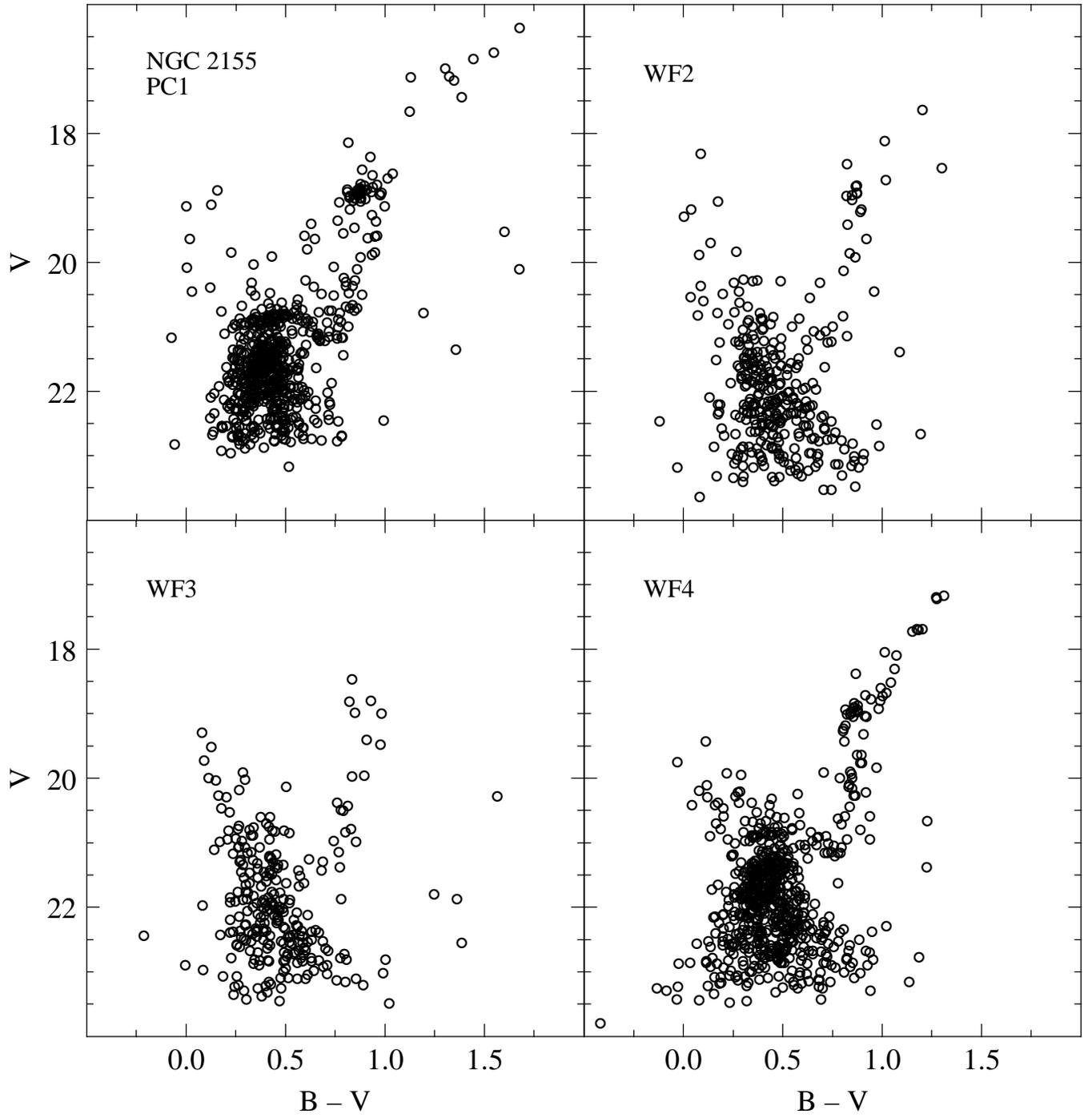
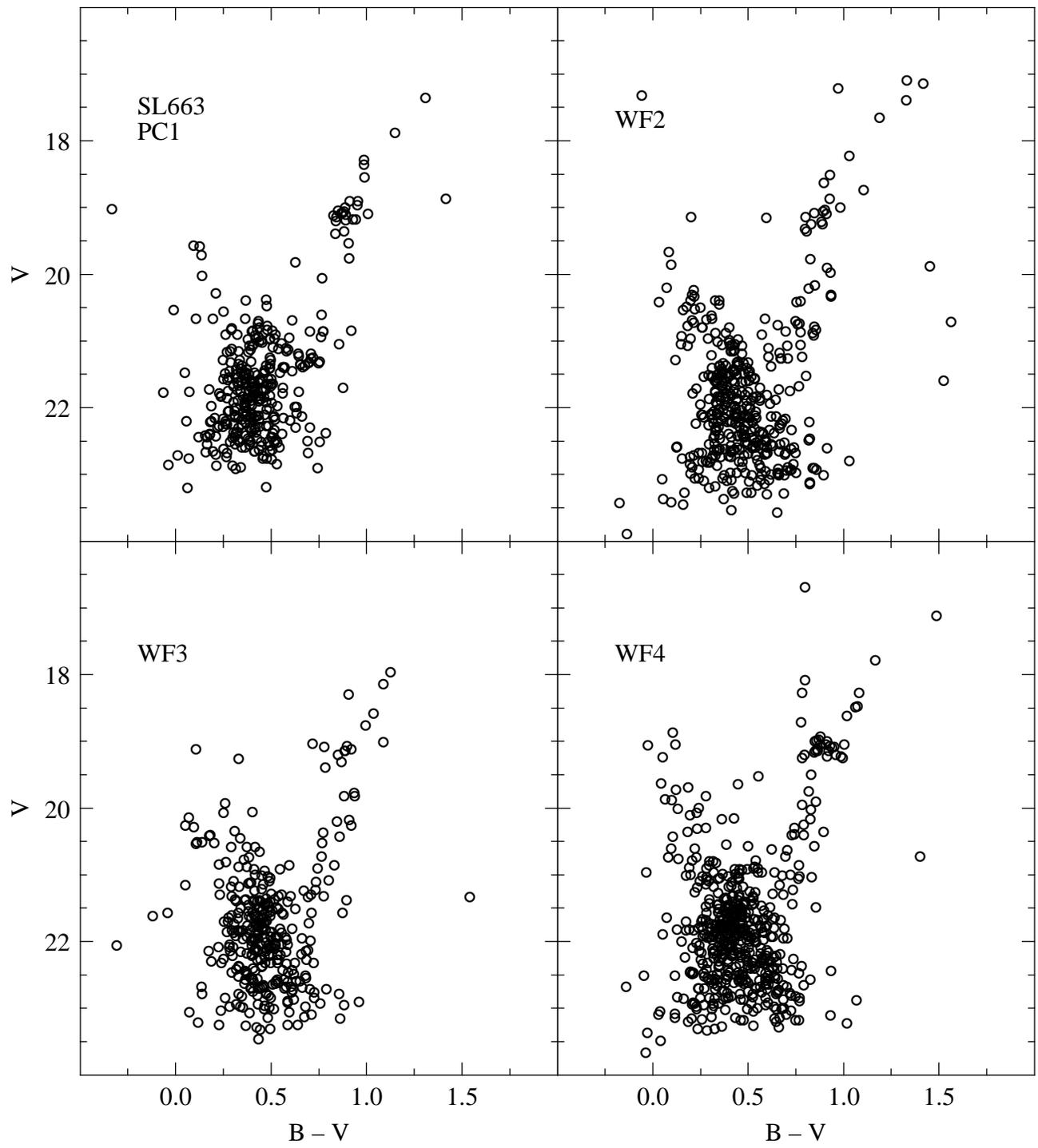
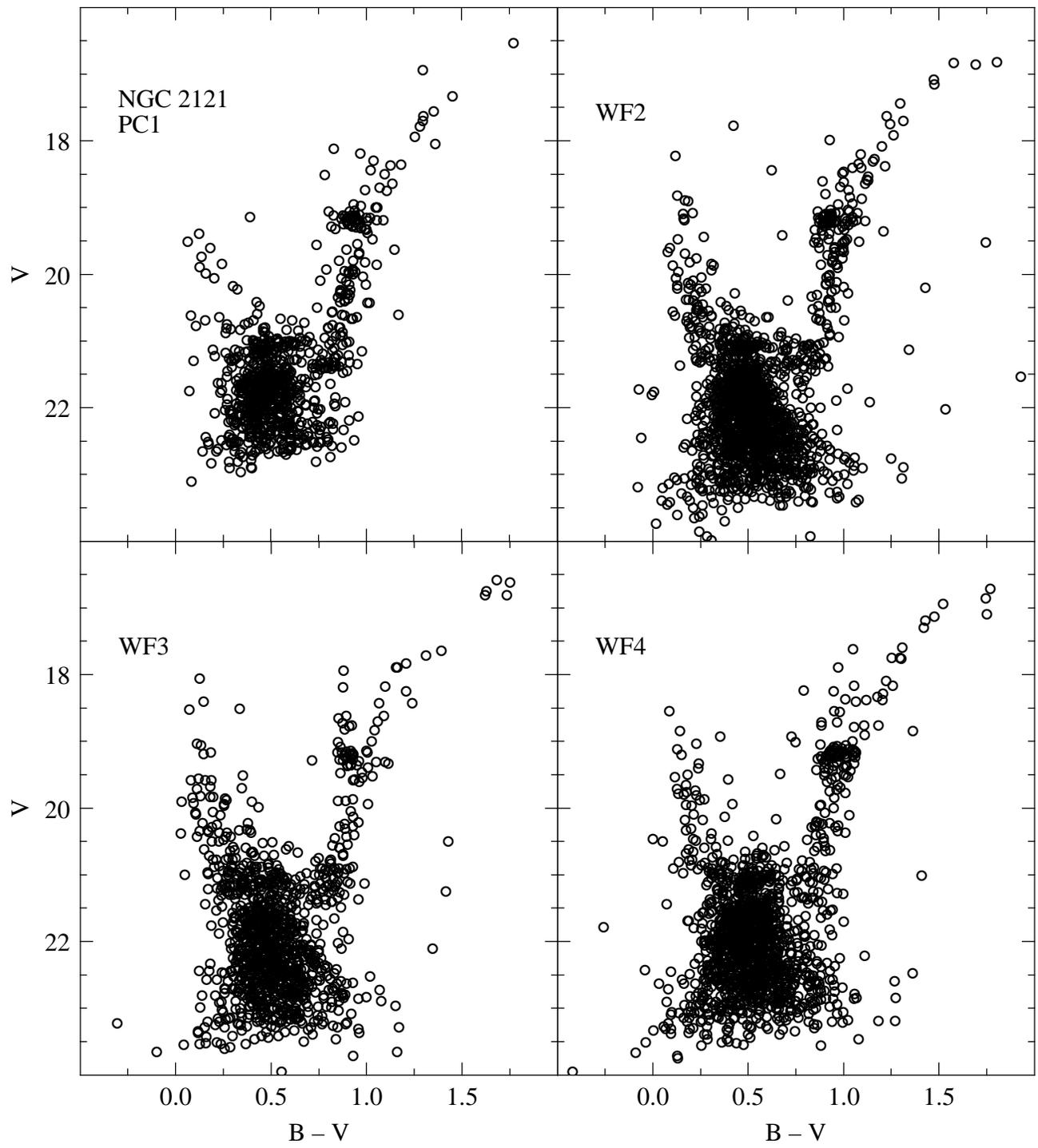
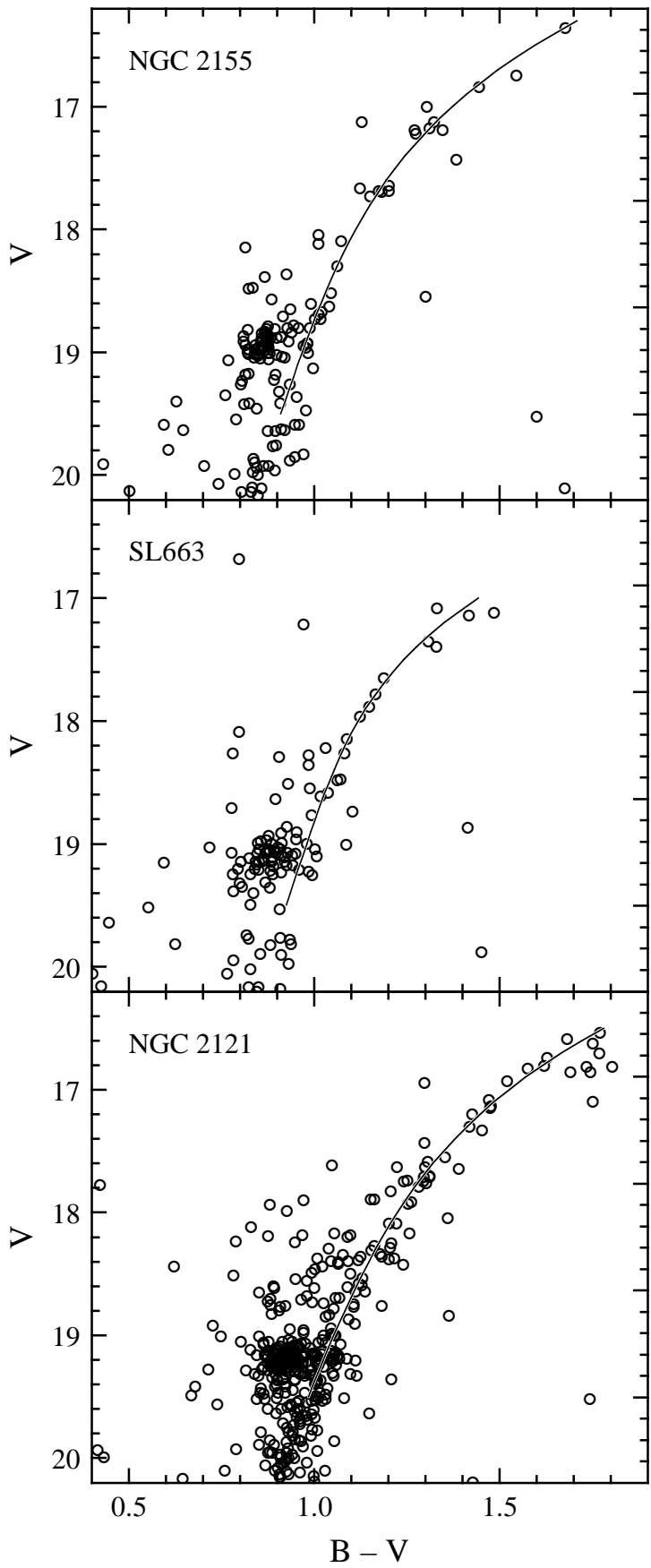
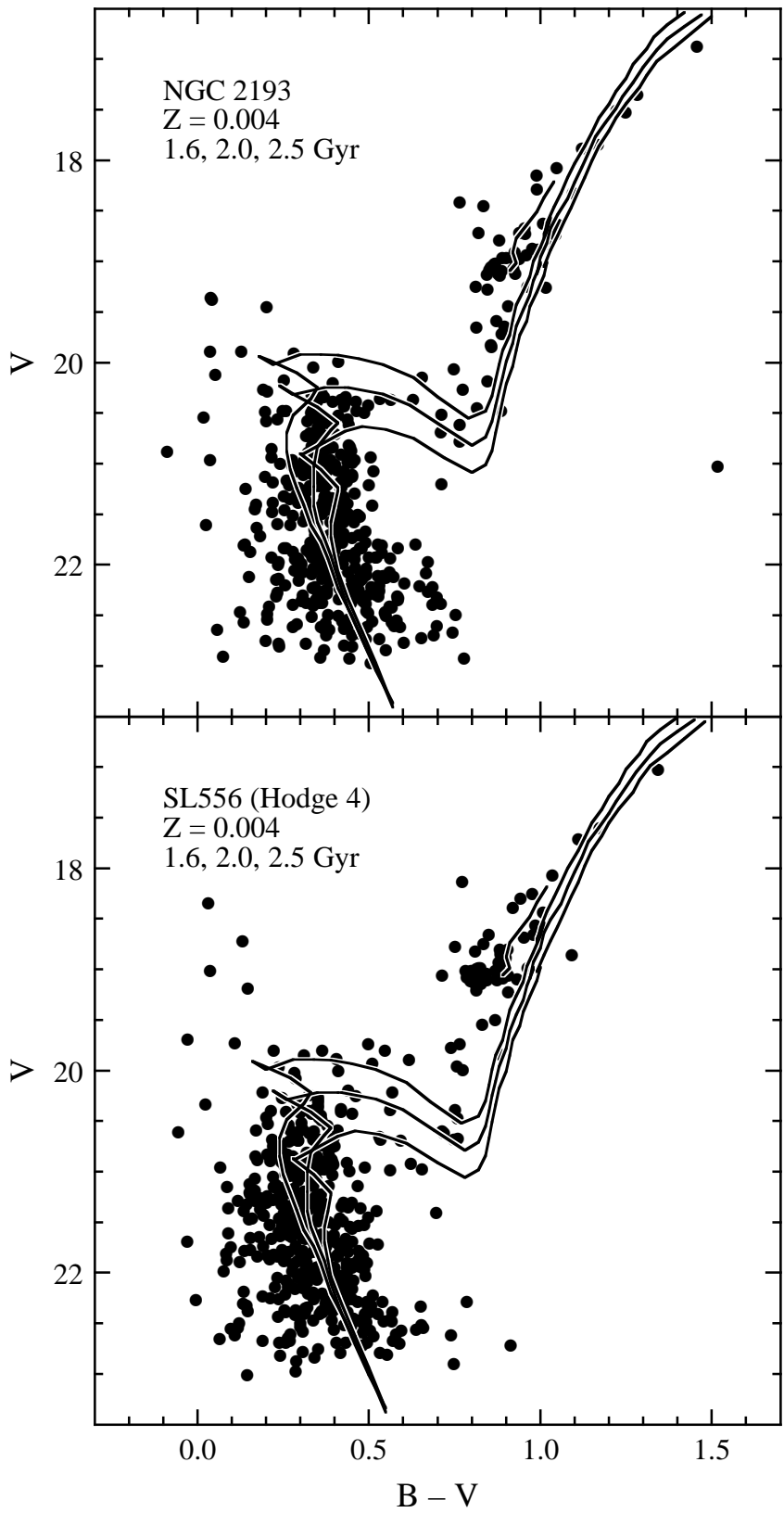
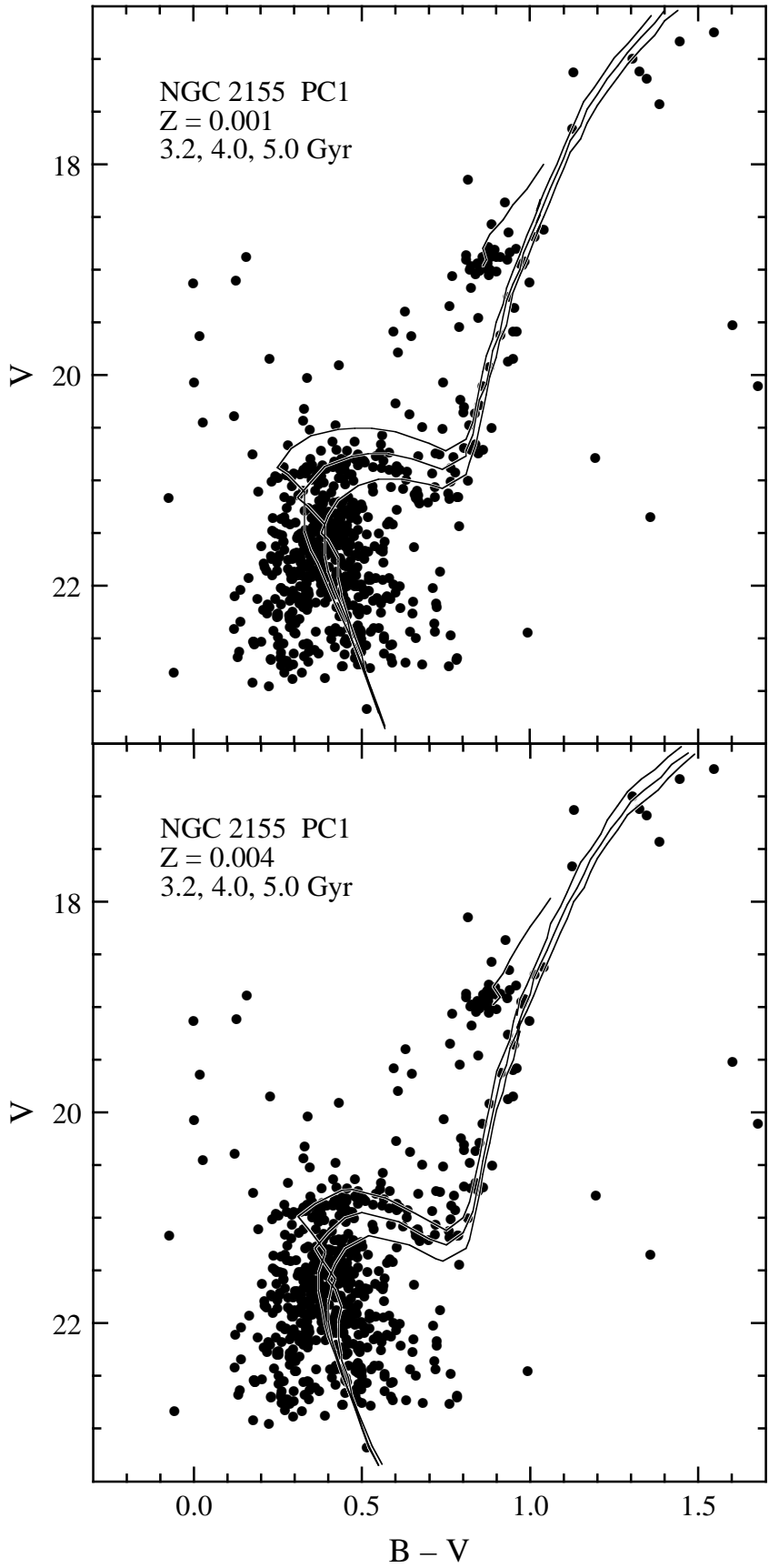
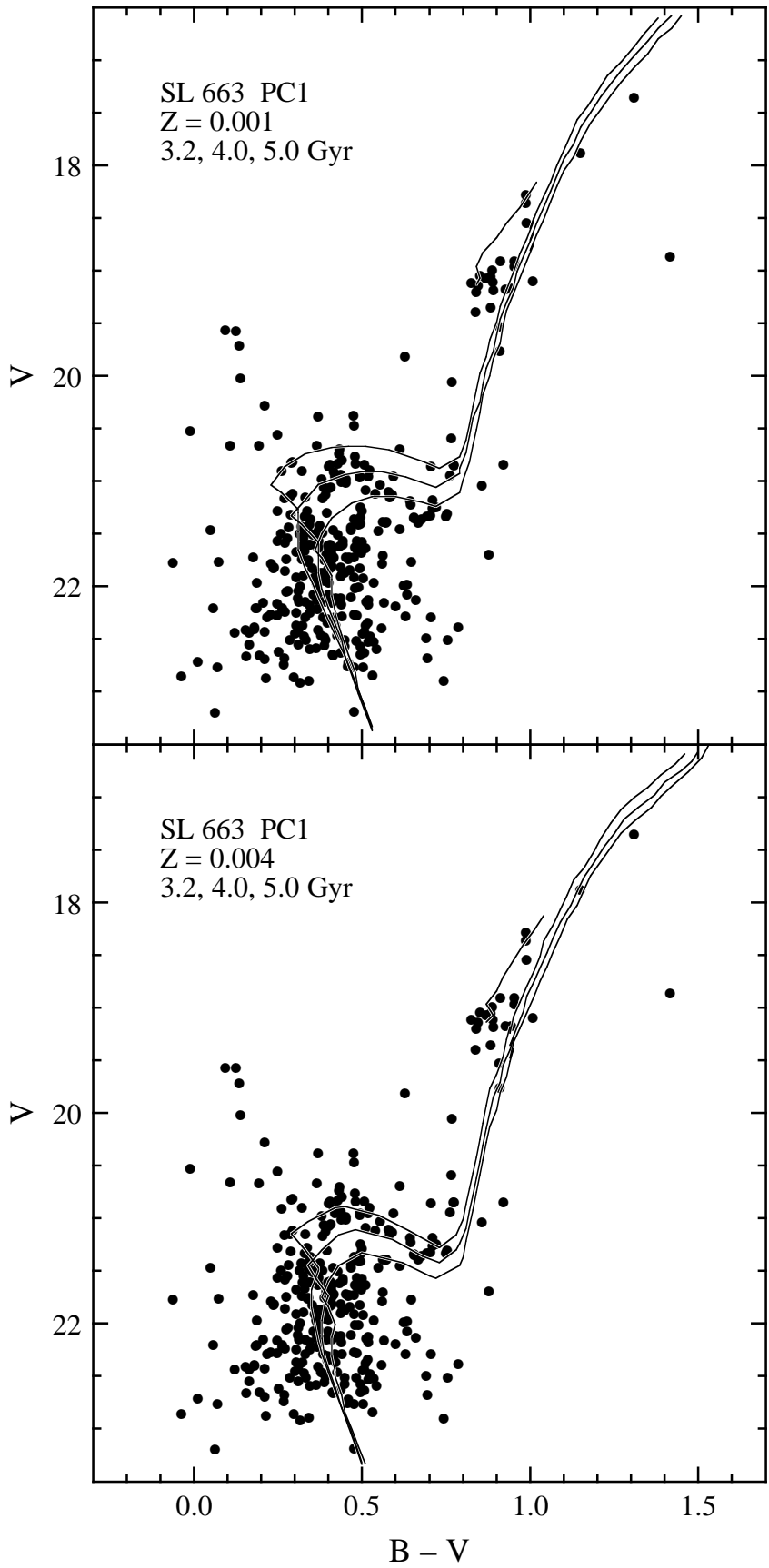
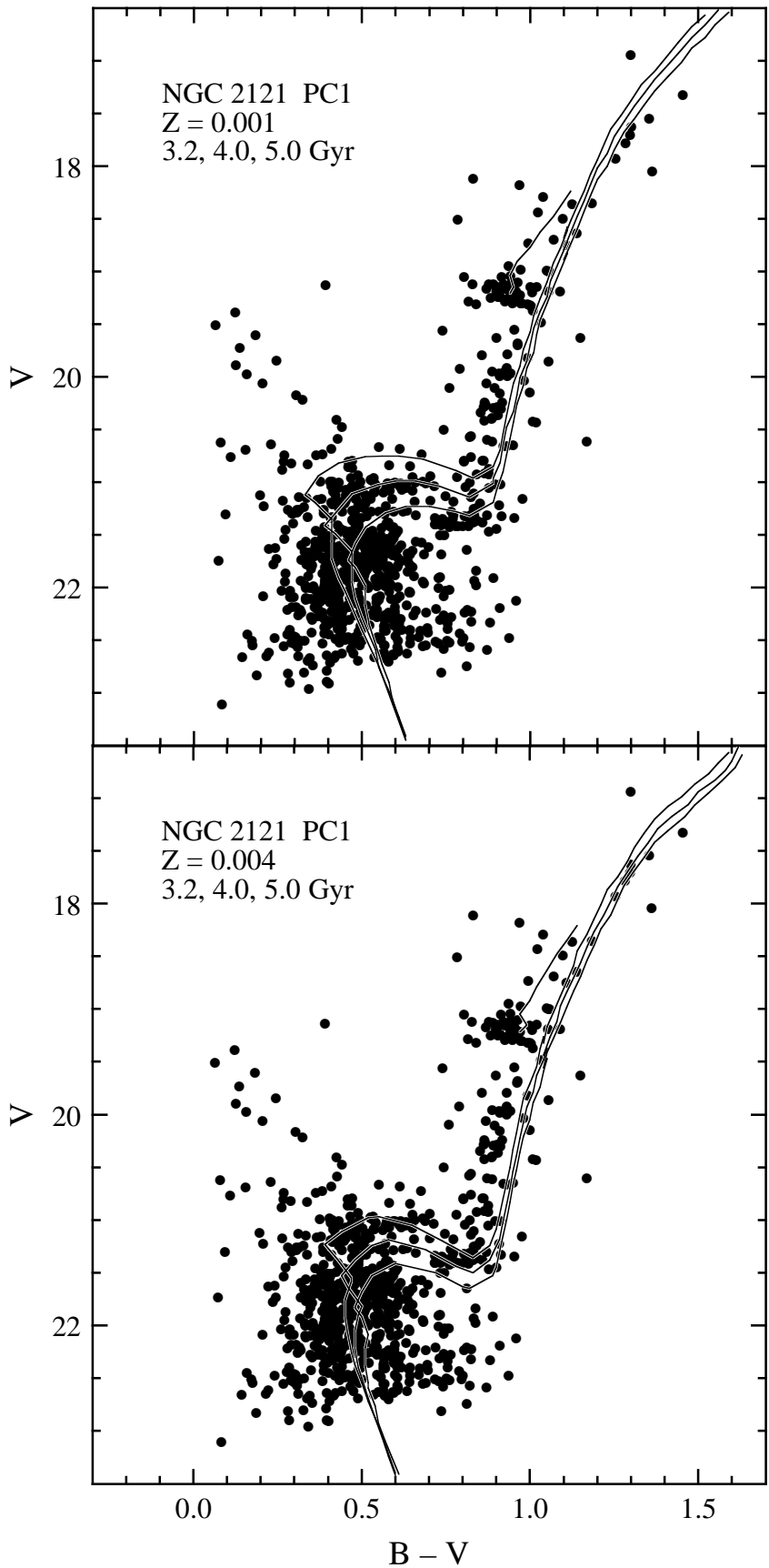
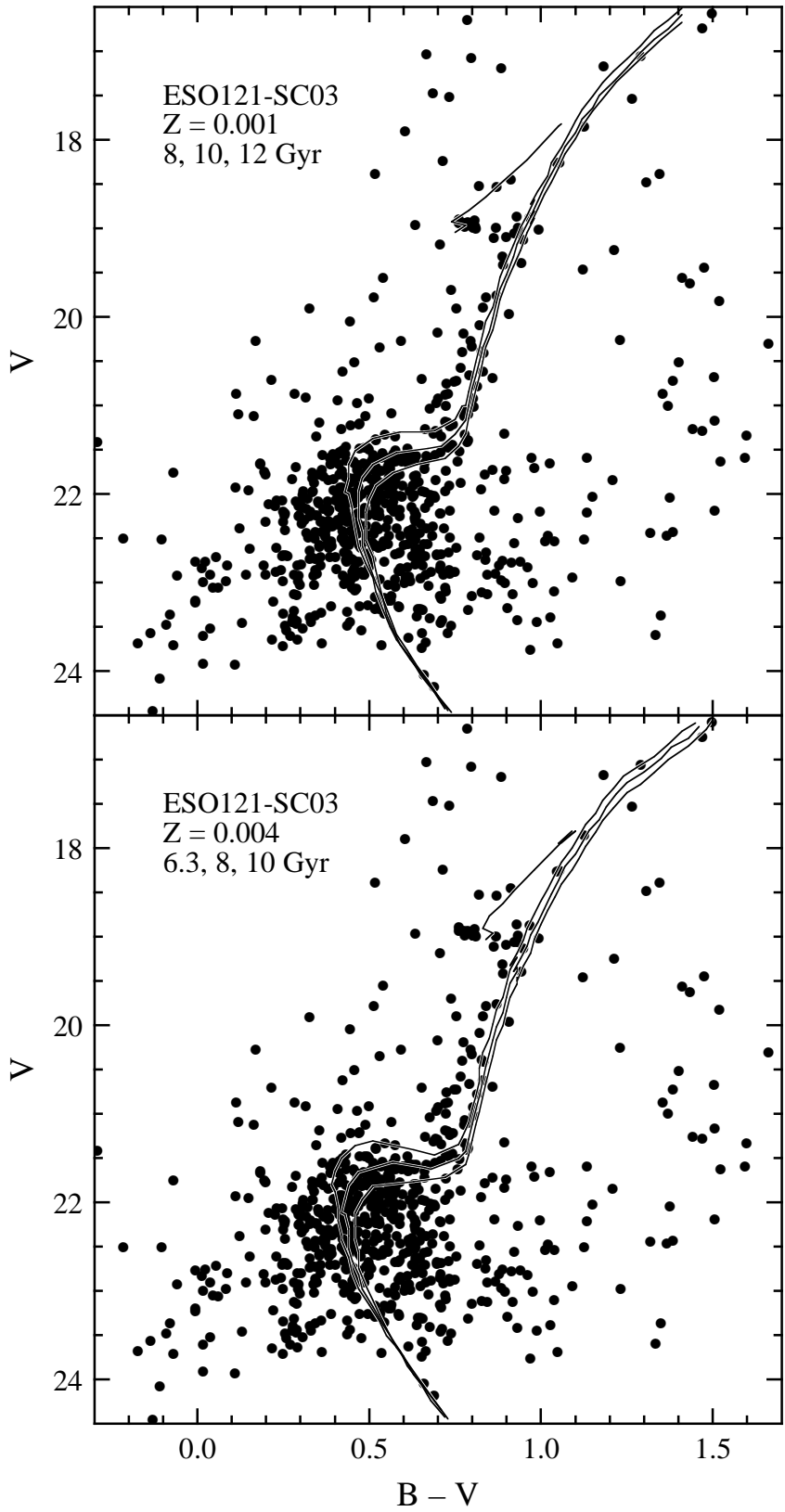
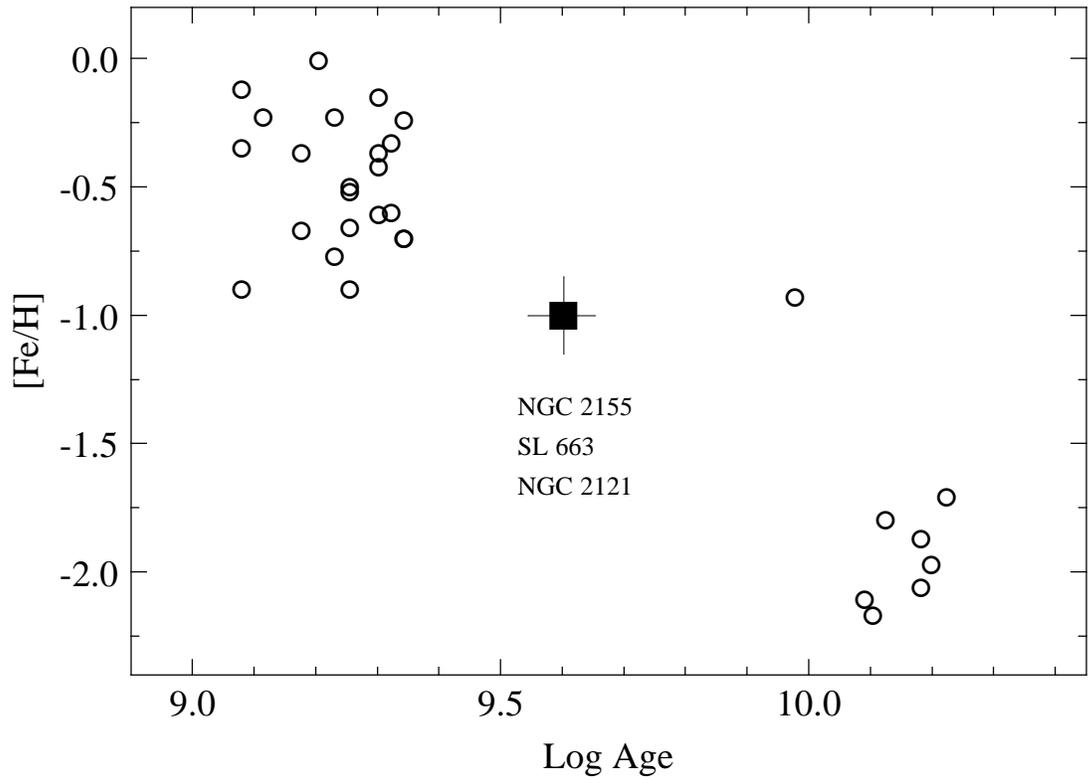
\end{document}